\documentclass[conference,preprint]{IEEEtran}
\IEEEoverridecommandlockouts    

\usepackage{multirow}
\usepackage{lipsum}
\usepackage{amsmath}
\usepackage{amssymb}
\usepackage[ruled,vlined]{algorithm2e}
\usepackage{graphicx}
\graphicspath{{./Figures/}}

\usepackage{acro} 
\usepackage{cite}
\usepackage{tikz}
\usepackage{textcomp}
\usepackage{hyperref}
\usepackage{lipsum}
\usepackage{siunitx}
\DeclareAcronym{BEV}{
  short = BEV,
  long = battery electric vehicle,
  long-plural-form = battery electric vehicles,
}

\DeclareAcronym{ADAS}{
  short = ADAS,
  short-plural-form = ADAS,
  long = Advanced Driver Assistance System,
  long-plural-form = Advanced Driver Assistance Systems,
}

\DeclareAcronym{ACC}{
  short = ACC,
  long = Adaptive Cruise Control,
  long-plural-form = N/A
}

\DeclareAcronym{UCC}{
  short = UCC,
  long = Urban Cruise Control,
  long-plural-form = N/A
}

\DeclareAcronym{UEQ}{
  short = UEQ,
  long = User Experience Questionnaire,
  long-plural-form = N/A
}

\DeclareAcronym{HMI}{
  short = HMI,
  long = Human Machine Interface,
  long-plural-form = N/A
}

\DeclareAcronym{LSA}{
  short = LKA,
  long = Lane Keeping Assistant,
  long-plural-form = N/A
}

\DeclareAcronym{CID}{
  short = CID,
  long = Central Information Display,
  long-plural-form = N/A
}

\DeclareAcronym{VR}{
  short = VR,
  long = Virtual Reality,
  long-plural-form = N/A
}

\DeclareAcronym{MISC}{
  short = MISC,
  long = Misery Scale,
  long-plural-form = N/A
}

\DeclareAcronym{NDRT}{
  short = NDRT,
  short-plural-form = NDRT,
  long = Non-Driving Related Task,
  long-plural-form = Non-Driving Related Tasks
}

\DeclareAcronym{VDL}{
  short = VDL,
  long = VanDerLaan,
  long-plural-form = N/A
}

\DeclareAcronym{DALI}{
  short = DALI,
  long = Driver-Activity-Load-Index,
  long-plural-form = N/A
}

\DeclareAcronym{DMS}{
  short = DMS,
  short-plural-form = DMS,
  long = Driver Monitoring System,
  long-plural-form = Driver Monitoring Systems
}
\newcommand{\meanstdinline}[2]{%
{\textit{M}~=~{#1},\allowbreak~\textit{SD}~=~{#2}}%
}

\newif\ifanonymize
\anonymizefalse 
\newcommand{\anonymtext}[2]{%
  \ifanonymize
    \textit{#2}%
  \else
    #1%
  \fi
}

\newcommand\copyrighttext{%
  \footnotesize \textcopyright 2026 IEEE. Personal use of this material is permitted.
  Permission from IEEE must be obtained for all other uses, in any current or future
  media, including reprinting/republishing this material for advertising or promotional
  purposes, creating new collective works, for resale or redistribution to servers or
  lists, or reuse of any copyrighted component of this work in other works. Paper accepted at IEEE ITSC 2026.}
\newcommand\copyrightnotice{%
\begin{tikzpicture}[remember picture,overlay]
\node[anchor=south,yshift=10pt] at (current page.south) {\fbox{\parbox{\dimexpr\textwidth-\fboxsep-\fboxrule\relax}{\copyrighttext}}};
\end{tikzpicture}%
}

\title{\LARGE \bf Investigating Driver Behavior in Complex Traffic Situations While Driving Partially Automated Vehicles}

\author{
	\parbox{\textwidth}{%
		\centering
		\anonymtext{Lukas Köning$^{1,2}$, Nataša Miličić$^{2}$, Klaus Bogenberger$^{1}$}{Anonymous Authors}%
	}%
    \thanks{This work involved human subjects or animals in its research. Approval of all ethical and experimental procedures and protocols was granted by \anonymtext{BMW's}{an anonymous} ethics committee, and performed in line with the Declaration of Helsinki.}%
	\anonymtext{\thanks{$^{1}$Chair of Traffic Engineering and Control, Technical University of Munich, 80333 Munich, Germany
		{\tt\small lukas.koening@tum.de, klaus.bogenberger@tum.de}}%
	\thanks{$^{2}$BMW AG, 80809 Munich, Germany
		{\tt\small natasa.milicic@bmw.de}}}{}%
}

\usepackage{geometry}
\begin{document}
	\newgeometry{top=0.75in,left=0.75in,
 right=0.75in,
 bottom=0.75in,}
	\maketitle
    \copyrightnotice
	\thispagestyle{empty}
	\pagestyle{empty}
	
	\begin{abstract}
		Traffic complexity critically influences driver task demands in partially automated vehicles, yet subjective perception and its behavioral indicators remain underexplored in real-world settings. This paper analyzes driver behavior - vehicle interaction, glance patterns, and guiding fixation - across varying levels of subjective traffic complexity, using real-world data from 20 drivers in real urban traffic.
        Traffic complexity was determined by expert labeling and served as ground truth for vehicle data. Statistical analysis of 16 driver behavior metrics revealed small but significant trends with increasing complexity: deviation from speed limit increased, brake rate increased while braking intensity decreased, horizontal gaze dispersion and entropy widened, and guiding fixation rate decreased, indicating defensive adaptation and perceptual shifts.
        Contributions include real-world validation of gaze metrics and guiding fixation under subjective complexity, novel insights from gaze and guiding fixation entropy metrics, and the identification of promising indicators~(driven speed, brake rate, gaze yaw entropy, guiding fixation rate) for complexity-adaptive partially automated vehicles.
        While based on a limited urban sample and expert-labeled subjective complexity, the findings provide a foundation for combined complexity scores and their integration into complexity-adaptive, partially automated vehicles, boosting human-like automation and enhancing safety and predictability in the traffic system.
	\end{abstract}
	
    \section{Introduction}
    \label{sec:introduction}
    Driving on a winding road could be a real pleasure in perfect weather with no traffic, or a demanding, complex task in foggy weather during rush hour. These differences in perceived subjective traffic complexity directly affect driver behavior, the demands placed on \ac{ADAS}, and traffic as a whole. 

    Modern partially automated \ac{ADAS} (SAE Level 2) take over lateral and longitudinal control, but the driver must still supervise the system~\cite{SAEJ3016}. 
    Maintaining driver engagement is challenging, as current systems often rely on punitive inattention warnings. Developing ADAS that promote active participation is thus a promising approach to enhancing road safety. 
    Understanding the current demand placed on the driver by the driving task is crucial to providing optimal driving assistance that promotes active participation and avoids overload.

    One important factor is the complexity of the current traffic situation, which strongly influences task demand, particularly for perception tasks \cite{Paxion2014}. Drivers tend to drive more defensively, reducing speed and increasing braking as complexity increases~\cite{Xie2021_2,Lyu2017}. Furthermore, increased complexity reduces lane-keeping performance and raises the probability of errors~\cite{Paxion2014}. 
    In the context of automated driving, research found declining take-over performance as complexity increases \cite{ScharfeScherf2021_2}.
    
    While many approaches quantify objective traffic complexity from environmental features, only a few studies examine driver behavior across levels of perceived subjective traffic complexity \cite{Yang2021}.
    Research has identified wider gaze behavior in more complex scenarios \cite{Kunst2025}. More concretely, a driving-simulator study reported increased vertical gaze dispersion but no change in horizontal gaze dispersion in more complex situations \cite{Halin2025}. Specifically for lane changes, more dispersed glances with increasing complexity were reported \cite{Wang2023}. These findings indicate that drivers increasingly need to gather information to support decision-making and situational awareness as complexity increases.
    In addition to conventional dispersion metrics, information-theoretic gaze measures, such as gaze entropy, have been proposed to characterize the complexity of visual scanning behavior~\cite{Goodridge2024}. However, these measures have not been systematically investigated with respect to subjective traffic complexity.

    To link a driver's gaze behavior to the driving task, the concept of guiding fixation can be used. It refers to the natural gaze behavior drivers use to guide the vehicle \cite{Lappi2022}. While driving, drivers focus their gaze primarily along the planned trajectory, typically about 2 seconds ahead \cite{Mole2021,Navarro2021}. This gaze behavior is defined as guiding fixation. Periodically, drivers also scan the traffic situation, including other road users and scenery. 

    Existing work is limited to specific maneuvers or simulator studies, lacks a subjective complexity ground truth, and rarely investigates partially automated vehicles. 
    Furthermore, the influence of traffic complexity on guiding fixation behavior and information-theoretic gaze measures has not been examined.  
    Therefore, this paper addresses the research question: 
    
    \textit{How does driver behavior - in terms of vehicle interaction, glance behavior, and guiding fixation - differ across traffic situations with varying levels of subjective traffic complexity?}

    This paper contributes to the state of the art by analyzing and validating driver behavior in real urban traffic with partially automated \ac{ADAS} as indicators of subjective traffic complexity, by extending existing gaze-based analyses with guiding fixation and information-theoretic gaze metrics across varying complexity, and by identifying driver behavior indicators that could be used in complexity-adaptive \ac{ADAS}.
    \restoregeometry
    \section{Data and Labeling}
	\label{sec:StudyAndLabeling}
	To investigate the effects of traffic complexity on drivers' behavior, driver-related data from a real-vehicle study in an urban environment in \anonymtext{Munich, Germany}{an anonymous Central European city}, were analyzed. The study is described in detail in the publication by Köning~et~al.~\cite{Koening2026}. As proxy for traffic complexity, a posteriori expert labeling of the drives was conducted. The following sections describe the study design, labeling framework, and results of the labeling.
	
	\subsection{Real-Vehicle Study}
    From the total of 57 participants in the study by Köning~et~al.~\cite{Koening2026}, a subsample of 20 participants (\meanstdinline{38.80 years}{10.24 years}, 3 female, 17 male) was selected to limit the labeling effort. The subsample was selected to reflect the diversity of the original sample as well as possible with respect to age, gender, driving experience, and previous experience with \ac{ADAS}. Every participant declared informed consent before participating in the study.

    The study was designed for within-subject analyses, with the level of shared longitudinal control (conventional vs. fully shared) as the within-subject factor. It consisted of a familiarization drive and two consecutive \ac{ADAS} drives, with the same route used for both \ac{ADAS} drives.
    For the analysis in this paper, only the first drives of participants were selected, during which the \ac{ADAS} was characterized by fully-shared longitudinal control, allowing the driver to brake without deactivating the \ac{ADAS} system \cite{Illgner2024,Koening2026}. Therefore, only segments with active \ac{ADAS} were analyzed, as fully-shared longitudinal control resulted in nearly 100\% \ac{ADAS} usage.
    Participants were informed in advance that they would use a partially automated \ac{ADAS} that assists with maintaining a safe distance from preceding vehicles and staying in lane.
    They were asked to use the \ac{ADAS} as much as possible, provided they felt comfortable driving with it.

    The route was designed to cover various urban driving scenarios in a Central European setting, as shown in Figure~\ref{fig:Route}. It spanned over 11\,km of multi-lane, single-lane, and side roads with a diverse set of traffic situations. 
    Additionally, the route passed multiple construction sites. 
    The study was conducted between 9 am and 5 pm in the summer, thereby allowing for varying traffic, lighting, and weather conditions.
    
    \ifanonymize
\else
    \begin{figure}[!htbp]
        \centering
        \includegraphics[width=0.40\textwidth, height=\textheight, keepaspectratio]{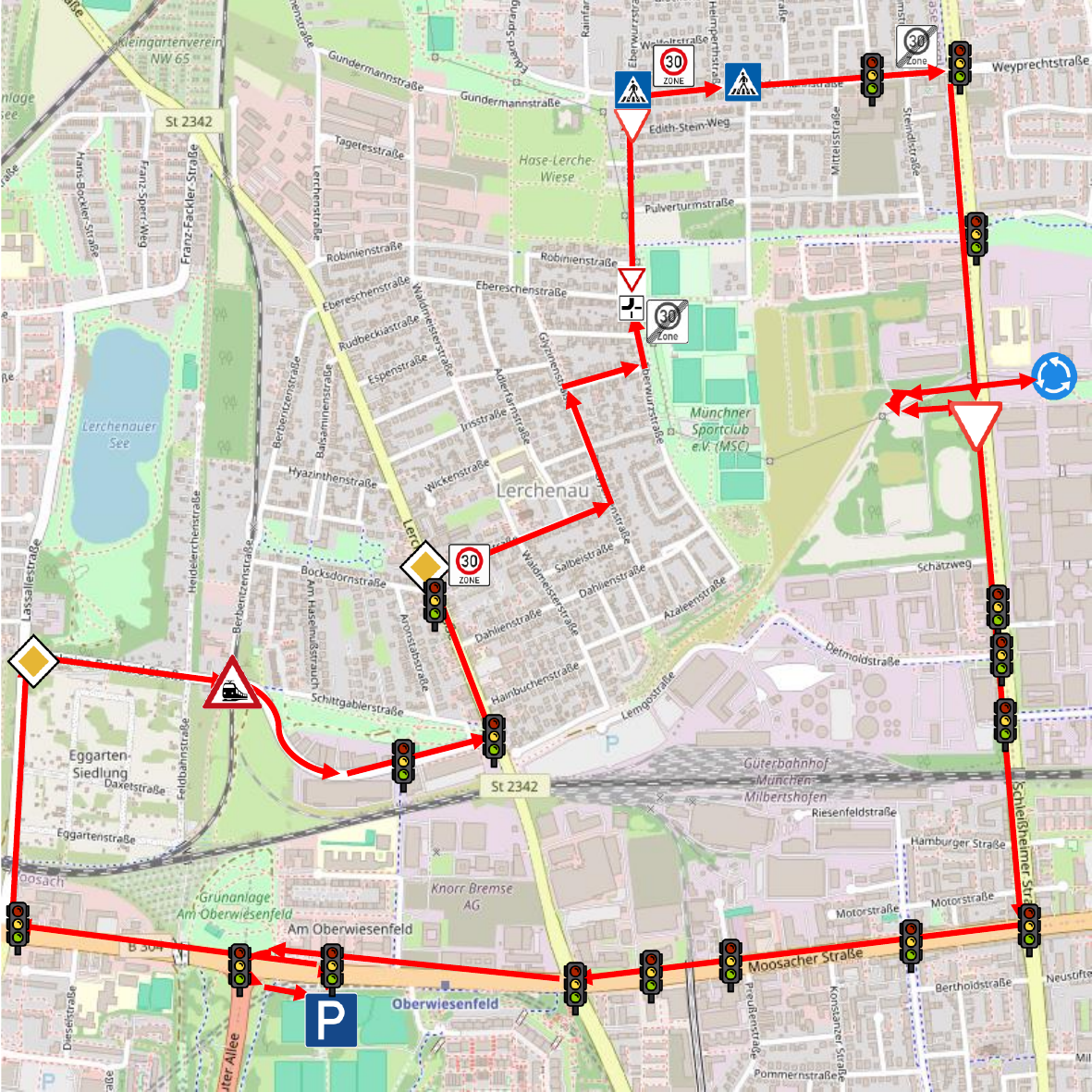}
        \caption{Route driven with the two \ac{ADAS} setups in the real-vehicle study.}
        \label{fig:Route}
    \end{figure}
    \fi

    For the real-vehicle study, \anonymtext{a pre-series BMW i5 M60}{an anonymous car} was used. The vehicle featured \anonymtext{the 2021 version of BMW's}{anonymized} \ac{ACC} and \ac{LSA}, classifying it as a partially automated \ac{ADAS}. To enable fully shared longitudinal control, a specialized software version was installed on the vehicle. Participants were informed that the specialized software carried a higher risk of system failures. Besides the specialized \ac{ACC} software, the vehicle's systems remained in their original production state.
    

    To analyze data from the vehicle's control units, a Python-based pipeline was developed. The extracted dataset includes GPS, speed, mileage, \ac{ADAS} activation, brake pedal usage, and driver monitoring data from the vehicle’s \ac{DMS}, located behind the steering wheel.

    \subsection{Traffic Complexity Labeling}
    In this study, subjective traffic complexity was assessed through expert evaluations of perceived situation complexity, serving as a proxy for drivers' experienced complexity. To establish expert-aggregated proxy labels for subsequent driver behavior analyses, experts conducted post hoc labeling of the drives. Three experts, covering the domains of automotive engineering, the psychology of human drivers, and usage safety, with at least three years of experience in their fields, were recruited. A posteriori labeling by the study participants was not possible due to logistical restrictions (post-study availability of all study participants). This a posteriori expert-labeling approach is widely used in continuous annotation tasks where post hoc participant labeling is not feasible and, when combined with a Dawid–Skene aggregation, provides a robust estimate of subjective traffic complexity \cite{Xie2019,Rastogi2022}.

    Three ordinal traffic complexity classes for traffic situations were defined:
    \begin{itemize}
        \item \textbf{simple}: Clearly structured; very few challenges and uncertainties; no adjustments necessary; lower level of attention needed.
        \item \textbf{normal}: Partially structured; few challenges and uncertainties; rare / minor adjustments necessary; normal level of attention needed.
        \item \textbf{complex}: Multifaceted / numerous variables; strong / frequent adjustments necessary; heightened level of attention needed.
    \end{itemize}
    Because experts should label the subjective traffic complexity of the shown traffic situation, the labeling task is inherently subjective.
    Following a one-hour session on coached labeling of one drive, the experts independently labeled the remaining 19 drives. A software tool that allowed for time-continuous labeling was provided. An example view is shown in Figure~\ref{fig:LabelingTool}. No predefined segments were provided, and the experts decided on their own when to switch between labels.

    \ifanonymize
\else
    \begin{figure}[!htbp]
        \centering
        \includegraphics[width=0.95\linewidth]{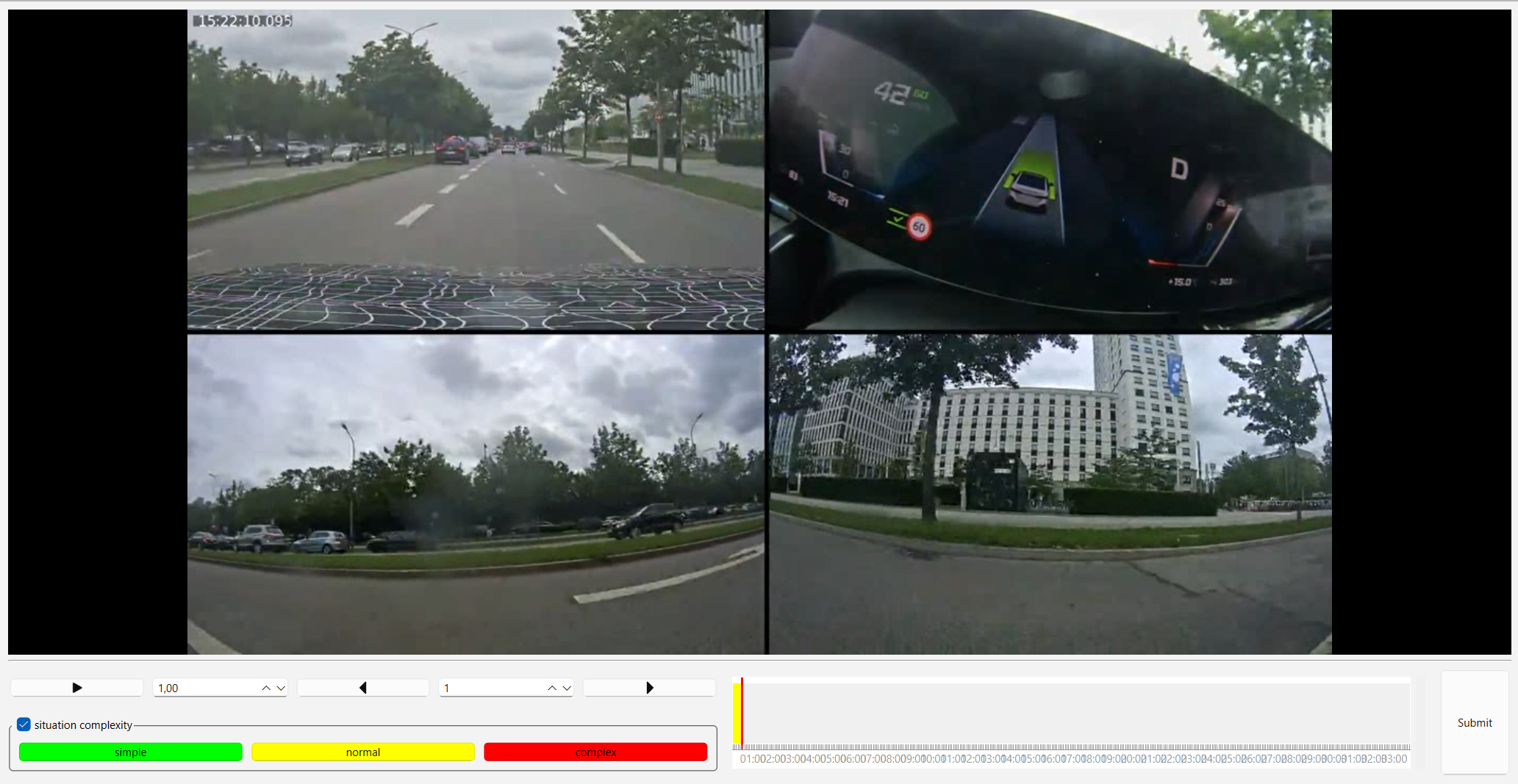}
        \caption{Labeling tool for expert labeling.}
        \label{fig:LabelingTool}
    \end{figure}
    \fi

    The labeling results from all experts were preprocessed to exclude segments with vehicle speeds below 1 km/h, since the analysis focuses on driver behavior while driving. Furthermore, a ±2-second tolerance window was applied to account for variations in label-change timestamps and expert reaction times \cite{Mariooryad2013,Mariooryad2015}. The individual labels were combined using the Dawid-Skene algorithm \cite{DawidSkene1979}.
    
    Overall, the class distribution of the combined labels is 27\% simple, 45\% normal, and 28\% complex. The inter-rater reliability results for the labeling are presented in Table~\ref{tab:Interraterreliability}. Cohen's~$\kappa$ and Krippendorff's~$\alpha$ were computed using a linear weighting to account for the ordinal scale of the complexity classes and to penalize high disagreement (simple vs. complex)~\cite{Vanbelle2024}. The results are interpreted according to the agreement strength by Landis and Koch~\cite{LandisKoch1977}. Krippendorff's~$\alpha$ of 0.57 indicates only moderate agreement due to the subjective labeling task and continuous, non-segmented video labeling~\cite{Mariooryad2013}. This is also reflected in the moderate pairwise inter-rater agreement and low Dawid-Skene rater weights.
    
    \begin{table}[htbp]
        \centering
        \begin{tabular}{cc}
            \textbf{Metric} & \textbf{Value}\\
            \hline
            Krippendorff's $\alpha$ & 0.5691\\
            Cohen's $\kappa$ (Expert 1-Expert 2) & 0.5225\\
            Cohen's $\kappa$ (Expert 1-Expert 3) & 0.5132\\
            Cohen's $\kappa$ (Expert 2-Expert 3) & 0.5900\\
            Cohen's $\kappa$ (Expert 1-Combined) & 0.6848\\
            Cohen's $\kappa$ (Expert 2-Combined) & 0.8189\\
            Cohen's $\kappa$ (Expert 3-Combined) & 0.7314\\
            Dawid-Skene rater weight Expert 1 & 0.2845\\
            Dawid-Skene rater weight Expert 2 & 0.3981\\
            Dawid-Skene rater weight Expert 3 & 0.1031\\
        \end{tabular}
        \caption{Results of inter-rater reliability analysis.}
        \label{tab:Interraterreliability}
    \end{table}
    
    However, the individual rater agreements with the combined labels, as computed by the Dawid-Skene algorithm, reach substantial values (0.68-0.82). Although inter-rater agreement is moderate, the strong agreement between individual raters and the combined labels justifies using the expert-aggregated proxy labels for later analyses. Nevertheless, a thorough sensitivity analysis of the vehicle speed filtering and tolerance window should be conducted in a follow-up publication. Furthermore, complementing in-situ driver feedback should be used to better define the traffic complexity.

    \section{Driver Behavior Analysis}
	\label{sec:DriverBehaviorAnalysis}
	To quantify the effects of traffic complexity on driver behavior, a diverse set of driver behavior metrics was analyzed. Building on the extracted dataset from the vehicle’s raw data, a Python-based pipeline was developed to preprocess the dataset, compute metrics, and perform statistical analyses using the expert labels.
    The preprocessing included filtering invalid data, sections where the vehicle speed was below 1 km/h, and sections where the driver was not looking outside the car’s windows, based on areas of interest determined by the vehicle’s \ac{DMS}. Furthermore, the data was linearly interpolated onto a fixed 50-ms grid to facilitate further calculations. Given the high-frequency sampling of the vehicle data during collection, ranging from 20 ms to 100 ms, the error resulting from the linear interpolation can be disregarded.
    In the following sections, the calculated behavioral metrics are defined, and the statistical results are presented. 
	
	\subsection{Behavioral Metrics}
	The evaluated behavioral metrics can be clustered into three categories: vehicle interaction, glance behavior, and guiding fixation behavior. To reduce the number of metrics presented, only the most relevant and technically feasible metrics, based on data availability, were selected. 

    \subsubsection{Vehicle Interaction}
    Regarding vehicle interaction, three signals were analyzed. First, the deviation of the driving speed from the current speed limit $\Delta v$ was evaluated by
    \begin{equation}
        \Delta v = \frac{v - v_{limit}}{v_{limit}}
    \end{equation}
    where $v$ is the driving speed measured by the vehicle’s own sensor and $v_{limit}$ is the current speed limit extracted from OpenStreetMap.

    Second, the driver’s brake interaction, recorded as a Boolean value, was evaluated. Based on the Boolean flags, time- and distance-based brake rates were calculated as
    \begin{equation}
        BR_{y} = \frac{y_{braking}}{y_{tot}}
    \end{equation}
    where $y_{braking}$ is either the time (t) or distance (s) driven while the driver used the brake, and $y_{tot}$ is the total time or distance driven within the rolling window of 15 seconds or 500 meters.

    Third, the acceleration during braking by the driver was analyzed using the actual brake acceleration $a_{braking}$, the standard deviation of the brake acceleration $SD(a_{braking})$, and the rolling Shannon entropy of the brake acceleration $H(a_{braking})$ using 30 fixed bins. The time window for rolling entropy calculations in this paper was set to 3 seconds.

    \subsubsection{Glance Behavior}
    To evaluate drivers' glance behavior, gaze yaw $\Psi_{g}$ and gaze pitch $\theta_{g}$ were extracted. The angles are defined in the east, north, up (ENU) coordinate system, where the x-axis of the right-handed coordinate system extends along the vehicle's longitudinal axis in the driving direction. For each signal $x$, the standard deviation $SD_s(x)$ and the rolling Shannon entropy $H_s(x)$ were computed. For the rolling Shannon entropy, 30 fixed bins were used. 
    
    To account for the nonlinearity of the spherical coordinate system, the standard deviations were computed as spherical standard deviations using the Eigenvectors rather than the raw values \cite{Ley2017}. The standard deviation and variance range from 0 (perfect alignment) to 1 (uniform distribution).
    In addition to the directly measured signals, the combined standard deviation $SD(g_{comb})$ and rolling entropy $H(g_{comb})$ for gaze yaw $\Psi_{g}$ and pitch $\theta_{g}$ were computed. Here, the bins for the entropy were set to 36 for gaze yaw and 18 for gaze pitch.
    
    Lastly, the fixation frequency $f_{fix}$ was calculated. A fixation change was defined as a combined change in gaze yaw $\Psi_{g}$ and pitch $\theta_{g}$ of at least 5° (accuracy of the installed \ac{DMS}). The fixation frequency was evaluated at each time step using a 15-second rolling window.

    \subsubsection{Guiding Fixation Behavior}
    Guiding fixation behavior integrates gaze behavior with the driving task. Here, a guiding fixation is defined when the driver's gaze yaw $\Psi_{g}$ lies within 10° of the target guiding fixation yaw, which is calculated as the yaw toward the point on the trajectory at a defined look-ahead distance. The look-ahead distance is defined by a look-up table of the average look-ahead distances for all participants in the study described in Köning et al. \cite{Koening2026} at specific speeds, as shown in Table \ref{tab:LookUpTableGF}. The driving speed was limited by the route's highest legal speed limit of 60 km/h.
    
    To evaluate the influence of traffic complexity on driver behavior related to the driving task, the guiding fixation behavior is analyzed using the guiding fixation rate $GFR$, calculated as
    \begin{equation}
        GFR = \frac{t_{GuidingFix}}{t_{tot}}
    \end{equation}
    where $t_{GuidingFix}$ is the time during which the driver exhibits guiding fixation behavior and $t_{tot}$ is the length of the rolling window. Here, a 15-second rolling window is used to calculate the guiding fixation rate.

    In addition to the guiding fixation rate $GFR$, the angular deviation $\delta_{GF}$ from the target guiding fixation is calculated as the dot product between the real gaze vector $v_{g}$ and the guiding fixation vector $v_{gf}$ as
    \begin{equation}
        \delta_{GF} = \arccos \left( \frac{\mathbf{v_{g}} \cdot \mathbf{v_{gf}}}{\|\mathbf{v_{g}}\| \, \|\mathbf{v_{gf}}\|} \right)
    \end{equation}
    where $\mathbf{v_{g}}$ consists of the gaze yaw $\Psi_{g}$ and gaze pitch $\theta_{g}$, and $\mathbf{v_{gf}}$ consists of the guiding fixation yaw and the pitch calculated from the eye position and the look-ahead distance. The standard deviation $SD(\delta_{GF})$ and the rolling Shannon entropy $H(\delta_{GF})$ were also evaluated using 30 bins for the entropy calculation. 
    \begin{table}
        \centering
        \begin{tabular}{cc}
            \textbf{Vehicle speed} & \textbf{Average look-ahead distance}\\
            \hline
            1-10 km/h & 11.8 m\\
            10-20 km/h & 14.5 m\\
            20-30 km/h & 15.0 m\\
            30-40 km/h & 15.3 m\\
            40-50 km/h & 16.0 m\\
            50-60 km/h & 16.5 m\\
        \end{tabular}
        \caption{Look up table for target look-ahead distance for the guiding fixation calculation.} %
        \label{tab:LookUpTableGF} %
    \end{table} %
    \subsection{Statistical Analysis}
    The statistical analyses were performed using the Kruskal-Wallis test as all data failed the Shapiro-Wilk test for normality. The post-hoc tests were performed using Bonferroni-corrected Dunn tests.
    A Bonferroni-corrected significance level of p\,$\textless$\,.003 (0.05 / 16 metrics) was set for all analyses to account for multiple significance tests. The effect size was determined using eta-squared~$\eta^2$ for Kruskal-Wallis tests and rank-biserial correlation coefficient $r_{rb}$ for post-hoc tests. To categorize the effect size as small, medium, or large, Cohen's guidelines for eta-squared~$\eta^2 =$ .01, .06, and .14 and for correlation coefficient~$r_{rb} =$ .10, .30, and .50 were used \cite{Cohen1988}.

    The results of the Kruskal-Wallis test across all evaluated metrics are summarized in Table \ref{tab:results_KruskalWallis}. The results of the Bonferroni-corrected post-hoc Dunn tests for metrics with at least small effect sizes in the Kruskal-Wallis test are shown in Table \ref{tab:results_PostHoc}.

    \begin{table*}[htb]
        \centering
        \centering
        \begin{tabular}{cccccccccc}
            \multirow{2}{5em}{\textbf{Metric}} & \multicolumn{2}{c}{\textbf{simple}} & \multicolumn{2}{c}{\textbf{normal}} & \multicolumn{2}{c}{\textbf{complex}} & \multirow{2}{5em}{\textbf{Effect Size}} & \multirow{2}{6em}{\textbf{Test Statistic}} & \multirow{2}{4em}{\textbf{p-Value}}\\
            & \textit{M} & \textit{SD} & \textit{M} & \textit{SD} & \textit{M} & \textit{SD} &  &  &\\
            \hline
            Speed Deviation $\Delta v\,[\si{\percent}]$ & -19.46 & 28.43 & -29.59 & 30.91 & -33.01 & 30.91 & 0.030 (s) & 4236.69 & \textless\,.001\\
            Brake rate (time) $BR_{t}$\,[\si{\percent}] & 2.792 & 10.64 & 4.861 & 10.96 & 11.01 & 17.48 & 0.057 (s) & 31675.35 & \textless\,.001\\
            Brake rate (distance) $BR_{s}$\,[\si{\percent}] & 3.827 & 6.154 & 5.165 & 6.464 & 6.919 & 9.201 & 0.025 (s) & 17102.46 & \textless\,.001\\
            \textbf{Brake acceleration} & & & & & & & & &\\
            $a_{braking}\,[\si{\meter/\second^2}]$ & -1.35 & 0.91 & -1.25 & 0.76 & -1.05 & 0.71 & 0.022 (s) & 561.57 & \textless\,.001\\
            $SD(a_{braking})\,[\si{\meter/\second^2}]$ & 0.34 & 0.28 & 0.38 & 0.26 & 0.32 & 0.25 & 0.010 (s) & 403.31 & \textless\,.001\\
            $H(a_{braking})\,[-]$ & 2.17 & 1.01 & 2.28 & 0.92 & 2.19 & 0.92 & 0.003 (n) & 82.09 & \textless\,.001\\
            \hline
            \textbf{Driver gaze yaw} & & & & & & & & &\\
            $SD(\Psi_{g})\,[-]$ & 0.140 & 0.030 & 0.146 & 0.025 & 0.148 & 0.025 & 0.015 (s) & 5845.92 & \textless\,.001\\
            $H(\Psi_{g})\,[-]$ & 1.539 & 0.796 & 1.759 & 0.814 & 1.776 & 0.826 & 0.016 (s) & 6357.06 & \textless\,.001\\
            \textbf{Driver gaze pitch} & & & & & & & & &\\
            $SD(\theta_{g})\,[-]$ & 0.046 & 0.013 & 0.047 & 0.012 & 0.047 & 0.012 & 0.001 (n) & 778.81 & \textless\,.001\\
            $H(\theta_{g})\,[-]$ & 1.829 & 0.598 & 1.845 & 0.589 & 1.842 & 0.606 & \textless\,0.001 (n) & 16.84 & \textless\,.001\\
            \textbf{Combined driver gaze} & & & & & & & & &\\
            $SD(g_{comb})\,[-]$ & 0.146 & 0.029 & 0.153 & 0.024 & 0.154 & 0.024 & 0.016 (s) & 5570.54 & \textless\,.001\\
            $H(g_{comb})\,[-]$ & 2.513 & 0.798 & 2.702 & 0.817 & 2.701 & 0.827 & 0.011 (s) & 4488.85 & \textless\,.001\\
            \textbf{Fixation frequency} & & & & & & & & &\\
            $f_{fix}\,[\si{\hertz}]$ & 2.486 & 1.484 & 2.816 & 1.366 & 2.711 & 1.361 & 0.010 (s) & 5669.21 & \textless\,.001\\
            \hline
            Guiding fixation rate $GFR$\,[\si{\percent}] & 76.92 & 22.90 & 73.16 & 22.59 & 66.11 & 25.10 & 0.029 (s) & 14853.72 & \textless\,.001\\
            \textbf{Angular deviation} & & & & & & & & &\\
            $\delta_{GF}\,[\si{\degree}]$ & 9.339 & 13.73 & 10.90 & 14.64 & 11.58 & 14.79 & 0.004 (n) & 4268.96 & \textless\,.001\\
            $SD(\delta_{GF})\,[\si{\degree}]$ & 6.206 & 6.370 & 6.893 & 6.476 & 7.011 & 6.304 & 0.003 (n) & 2461.77 & \textless\,.001\\
            $H(\delta_{GF})\,[-]$ & 1.144 & 0.750 & 1.335 & 0.773 & 1.409 & 0.782 & 0.018 (s) & 7212.56 & \textless\,.001\\
        \end{tabular}
        \caption{Results of the statistical analysis of driver behavior metrics. Effect size $\eta^2$ interpreted as negligible (n), small (s), medium (m), large (l) according to the previous definition.}
        \label{tab:results_KruskalWallis}
    \end{table*}

    \begin{table*}[htb]
        \centering
        \begin{center}
        \begin{tabular}{ccccccc}
            \multirow{2}{5em}{\textbf{Metric}} & \multicolumn{2}{c}{\textbf{simple-normal}} & \multicolumn{2}{c}{\textbf{normal-complex}} & \multicolumn{2}{c}{\textbf{simple-complex}}\\
            & Effect Size & p-Value & Effect Size & p-Value & Effect Size & p-Value \\
            \hline
            Speed Deviation $\Delta v$\,[\si{\percent}] & -0.196 (s) & \textless\,.001 & -0.070 (n) & \textless\,.001 & -0.266 (s) & \textless\,.001\\
            Brake rate (time) $BR_{t}$\,[\si{\percent}] & 0.130 (s) & \textless\,.001 & 0.217 (s) & \textless\,.001 & 0.337 (m) & \textless\,.001 \\
            Brake rate (distance) $BR_{s}$\,[\si{\percent}] & 0.154 (s) & \textless\,.001 & 0.163 (s) & \textless\,.001 & 0.320 (m) & \textless\,.001 \\
            \textbf{Brake acceleration} & & & & & &\\
            $a_{braking}\,[\si{\meter/\second^2}]$ & 0.036 (n) & .003 & 0.163 (s) & \textless\,.001 & 0.191 (s) & \textless\,.001 \\
            $SD(a_{braking})\,[\si{\meter/\second^2}]$ & 0.142 (s) & \textless\,.001 & -0.141 (s) & \textless\,.001 & 0.011 (n) & 1.0 \\
            \hline
            \textbf{Driver gaze yaw} & & & & & &\\
            $SD(\Psi_{g})\,[-]$ & 0.144 (s) & \textless\,.001 & 0.020 (n) & \textless\,.001 & 0.168 (s) & \textless\,.001 \\
            $H(\Psi_{g})\,[-]$ & 0.156 (s) & \textless\,.001 & 0.013 (n) & \textless\,.001 & 0.167 (s) & \textless\,.001 \\
            \textbf{Combined driver gaze} & & & & & &\\
            $SD(g_{comb})\,[-]$ & 0.143 (s) & \textless\,.001 & 0.014 (n) & \textless\,.001 & 0.161 (s) & \textless\,.001 \\
            $H(g_{comb})\,[-]$ & 0.135 (s) & \textless\,.001 & -0.001 (n) & \textless\,.001 & 0.134 (s) & \textless\,.001 \\
            \textbf{Fixation frequency} & & & & & &\\
            $f_{fix}\,[\si{\hertz}]$ & 0.162 (s) & \textless\,.001 & -0.050 (n) & \textless\,.001 & 0.113 (s) & \textless\,.001 \\
            \hline
            Guiding fixation rate $GFR$\,[\si{\percent}] & -0.138 (s) & \textless\,.001 & -0.173 (s) & \textless\,.001 & -0.294 (s) & \textless\,.001 \\
            \textbf{Angular deviation} & & & & & &\\
            $H(\delta_{GF})\,[-]$ & 0.145 (s) & \textless\,.001 & 0.055 (n) & \textless\,.001 & 0.197 (s) & \textless\,.001 \\
        \end{tabular}
        \end{center}
        \caption{Results of the post-hoc test of the driver behavior metrics where Kruskal-Wallis revealed at least small effect sizes. Effect size $r_{rb}$ of post-hoc tests interpreted as negligible (n), small (s), medium (m), large (l) according to the previous definition.}
        \label{tab:results_PostHoc}
    \end{table*}
    
    \subsubsection{Vehicle Interaction}
    The interaction between the driver and the vehicle was analyzed with respect to deviation of driving speed from the speed limit, brake rate, and brake acceleration. 
    The deviation from the speed limit increases significantly with increasing complexity and is negative at all complexity levels. The post-hoc tests show significant, small effects across all complexity levels, except for the effect size between normal and complex, which is negligible.
    Both time- and distance-based brake rates increase significantly with increasing complexity, and post-hoc tests revealed significant differences across all complexity levels. The effect sizes are small for all effects.
    Brake acceleration, which indicates how hard drivers were braking, shows the inverse pattern. As complexity increases, drivers brake less hard. The post-hoc tests revealed small effect sizes, except for the comparison between simple and normal, which was not significant and showed a negligible effect size. The standard deviation of brake acceleration exhibits an inverse-U shape, with a small, significant effect. 
    The entropy of brake acceleration follows the same pattern as the standard deviation, but the significant effect has a negligible effect size. 

    \subsubsection{Glance Behavior}
    The driver’s glance behavior was examined in terms of gaze yaw, gaze pitch, combined gaze, and fixation frequency. 
    The statistical analysis reveals a slight but significant increase in the standard deviation and entropy of gaze yaw with increasing traffic complexity. Post-hoc tests show small effect sizes for comparisons between simple and the other complexity classes. However, the comparison between normal and complex shows only a negligible effect. Descriptively, the results indicate low but increasing angular dispersion and unpredictability of gaze behavior. 
    Regarding gaze pitch, no descriptive differences in standard deviation or entropy are observed across the complexity classes. The effect sizes are negligible even though the differences are statistically significant.
    The statistical analysis of combined gaze revealed a pattern similar to that of gaze yaw alone, with increases in standard deviation and entropy as traffic complexity increased. The post-hoc tests also show the same pattern, with a negligible effect size for the comparison between normal and complex. 
    Lastly, the fixation rate results show an increase with increasing complexity. The post-hoc tests show small effect sizes, except for the comparison between normal and complex, which shows a negligible effect. Therefore, the descriptive inverse U-curve pattern cannot be shown to be significant, and only the increase remains interpretable.
    
    \subsubsection{Guiding Fixation Behavior}
    The guiding fixation behavior was analyzed in terms of the guiding fixation rate and angular dispersion from the guiding fixation point. The results show a statistically significant decrease in the guiding fixation rate with increasing traffic complexity, as confirmed by post hoc tests. The effect size is small.
    In addition, the angular dispersion shows a slight increase with increasing complexity. This pattern is also evident in the angular dispersion's standard deviation and entropy. However, only the entropy exhibits a small effect size. Here, the post-hoc tests revealed small effect sizes except for the comparison between normal and complex, which has a negligible effect size.

    \section{Discussion}
	\label{sec:Discussion}
    Data from a real-vehicle study were evaluated using 16 driver behavior-related metrics covering vehicle interaction, glance behavior, and guiding fixation behavior across three expert-labeled traffic complexity classes.

    The analyses revealed significant patterns in vehicle interactions, including greater deviation from the speed limit, higher brake rates, and lower braking intensity as complexity increased. 
    These effects reflect drivers' increased need to reduce situational dynamics, thereby increasing perceptual time as complexity rises \cite{Paxion2014}. The tendency toward more braking but less hard braking in complex situations suggests more defensive, anticipatory driving \cite{Xie2021_2,Lyu2017}. 
    Reduced driving speed is one expected primary behavioral response to increased traffic complexity, acting as a mediating variable on the causal path from complexity to other behavioral metrics, rather than as a confounder. This is because speed reduction is causally influenced by complexity rather than independent of it~\cite{Paxion2014}. Accordingly, the total effect of complexity on driver behavior, including both direct and speed-mediated pathways, is the relevant measure for this paper. Controlling for speed would remove ecologically valid adaptive responses and would address a more limited research question.
    

    Regarding glance behavior, both the standard deviation and the entropy of horizontal and combined gaze increase with significant differences between simple and other complexity classes. The pattern is consistent with a broader, less predictable horizontal scanning strategy, as drivers need to gather more environmental information as complexity increases.
    Fixation frequency increased from simple to normal situations and then remained at a comparable level in complex situations. Thus, it can be interpreted as generally higher in more complex situations, rather than as systematically differentiating between normal and complex situations.
    No significant patterns with noticeable effect sizes were observed for vertical gaze dispersion. 
    The results contrast with those of Halin et al. \cite{Halin2025}, who reported increased dispersion in vertical gaze but no change in horizontal gaze. This could stem from differences between simulated and real driving environments \cite{Galante2018}. The effects of situational demand seem to outweigh the effects of gaze concentration patterns during cognitive distraction in real driving scenarios \cite{Paxion2014}.
    Taken together, the increases in horizontal gaze dispersion and gaze entropy indicate not only greater gaze dispersion but also less predictable sampling of the scene as complexity increases, consistent with recent work using gaze entropy as a descriptor of scanning complexity in driving and other visuomotoric tasks \cite{Melnyk2024,Goodridge2024}.

    The influence on driving-task-related gaze behavior was evaluated by harnessing the concept of guiding fixation. The results revealed a decrease in the guiding fixation rate, accompanied by an increase in angular deviation entropy as complexity increased. This reflects a shift in drivers' attention away from guiding-related driving tasks toward perceiving the environment as complexity increases. 

    Overall, the results presented in this paper reflect a more anticipatory and defensive driving style, along with broader and more unpredictable gaze patterns that shift away from the pure guiding task as traffic complexity increases. These behavioral shifts appear to be driven by a greater need to mitigate situational uncertainty, as greater complexity compels drivers to trade off speed for perceptual overhead. The results align with and extend findings by Kunst~et~al.~\cite{Kunst2025} and Wang~et~al.~\cite{Wang2023}, offering novel insights into new metrics and guiding fixation behavior. The most promising metrics that differ across traffic complexity are deviation from speed limit, brake rate, brake acceleration, horizontal gaze dispersion, and the guiding fixation rate. Changes in these metrics should be interpreted by \ac{ADAS} as indicators of traffic complexity and of changing external perceptual demands on the driver, rather than as signs of distraction or inattentiveness. Adapting \ac{ADAS} behavior to the traffic complexity perceived by the driver, e.g., by reducing speed in complex scenarios or increasing the time gap of \ac{ACC}, could lead to more human-like driving behavior from the \ac{ADAS}, benefiting the driver and surrounding traffic by improving the predictability of vehicle behavior and reducing implausible behavior during takeovers.

    The presented statistical results reveal significant differences but small effect sizes for the investigated metrics. Such small effects are typical in real-world driving studies, which often exhibit high variability \cite{Goodridge2024, Halin2025}. Statistical tests such as the Kruskal-Wallis test tend to detect even small significant differences in large data sets, as investigated in this paper \cite{Cohen1988}. Also, the Bonferroni correction had no effect as the differences remain highly significant. Therefore, the combination of effect size and significance was used to interpret the results. Another finding from the presented statistical results is that single behavior metrics cannot capture the variance in human behavior that depends on the complexity of the traffic situation. Hence, a combination of multiple metrics into one complexity score should be investigated.

    The results presented are based on a limited set of scenarios due to a sample of 20 individual drivers, unequal gender distribution, urban-only driving, and driving solely with active, partially automated \ac{ADAS}. Therefore, the analysis should be conducted on a more diverse dataset that includes additional drivers and environmental conditions.
    Another limitation concerns the traffic complexity labeling. While the original drivers were no longer available, the expert labels yielded only moderate inter-rater agreement. As a posteriori labeling lacks the real driving dynamics, situational aspects not covered by the video, and the real consequences of actions, it can only serve as an approximation of the real, felt subjective traffic complexity. Because these labels serve as a proxy for drivers’ perceived complexity, future studies should complement expert ratings with in-situ driver feedback. They should also aim to validate this labeling approach against objective measures to enhance generalizability.
    Furthermore, traffic complexity was labeled using three ordinal complexity classes. To further investigate the subjective effects of traffic complexity, more extensive labeling with a continuous complexity score should be considered. Also, because the label definition already included statements about needed adjustments, criterion overlap with the analyzed metrics cannot be ruled out.
    Lastly, the study focused only on attentive drivers. The effects of traffic complexity on inattentive drivers should also be examined, especially during transitions between attentiveness and inattentiveness.
    \section{Conclusions}
	\label{sec:conclusion}
	In conclusion, the results from the driver behavior analysis in real-world urban driving with different subjective traffic complexity levels revealed differences in vehicle interaction, glance behavior, and guiding fixation behavior. 
    As complexity increased, deviations from the speed limit increased, brake rate increased, and braking intensity decreased significantly, indicating more defensive driving. Regarding glance behavior, drivers showed increased horizontal and combined gaze dispersion and unpredictability as complexity increased. Also, fixation frequency increased overall. Lastly, guiding fixation behavior showed a decrease in guiding fixation rate and an increase in angular deviation entropy with increasing complexity, indicating that drivers prioritized environmental perception over guiding behavior. 

    The presented study advances the current state of the art by conducting the first investigation of driver behavior in real-world urban traffic involving partially automated vehicles, compared with subjective traffic complexity levels. Furthermore, the already investigated driver behavior metrics are validated in real traffic and extended by insights from guiding fixation behavior and information-theoretic measures. The results reveal that deviation from the speed limit, brake rate, brake intensity, horizontal gaze dispersion, and guiding fixation rate are exploratory but promising indicators of subjective traffic complexity.
    By adapting \ac{ADAS} to traffic complexity through complexity-adaptive human-machine interfaces, complexity-adaptive lateral and longitudinal control behaviors, or context-aware driver-monitoring systems, partially automated vehicles could better integrate the driver into the driving task and deliver more human-like driving, thereby improving the performance and safety of \ac{ADAS} and the traffic system as a whole.

    A follow-up paper will present the aforementioned sensitivity analysis of the labels and parameters used, and investigate additional driver behavior metrics, also on the participant level. Moreover, a combined complexity score will be developed and validated using principal component analysis, and correlations with objective variables of a traffic situation will be analyzed to bridge the gap between subjective and objective traffic complexity, as a basis for complexity-aware \ac{ADAS}. 
	
	
	\bibliographystyle{IEEEtran}
	\bibliography{root} 
	
\end{document}